\documentclass[ preprint,aps,nofootinbib,superscriptaddress,groupedaddress]{revtex4}

\usepackage{graphics}
\usepackage{makeidx}
\usepackage{epsfig}
\usepackage[caption=false]{subfig}
\usepackage{colortbl}
\usepackage{colordvi}
\usepackage{verbatim}
\usepackage{amsmath, amsthm}
\usepackage{enumerate}
\usepackage{wrapfig}
\usepackage{hyperref}
\usepackage{dcolumn}
\usepackage{bm}
\usepackage{bbm}
\usepackage{amssymb,mathrsfs}
\begin{document}

\title{ The time measurement problem in quantum cosmology}

\author{Nirmalya Kajuri}
\email{nirmalya@physics.iitm.ac.in}
 \affiliation{ Department of Physics, Indian Institute of Technology Madras,\\ Chennai 600036, India}

\begin{abstract}
In the canonical approach to quantization of gravity, one often uses relational clock variables and an interpretation in terms of conditional probabilities to overcome the problem of time. In this essay we show that these suffer from serious conceptual issues. 
\\
\vspace{10mm}

\textbf{Essay written for the Gravity Research Foundation 2017 Awards for Essays on Gravitation}

\end{abstract}
\maketitle
\section{Introduction}

The canonical route to quantum gravity is fraught with severe conceptual difficulties, of which the problem of time is certainly the most troublesome \cite{kuchar}. The problem is essentially that the Hamiltonian is a constraint. So physical observables, which by definition commute with all the constraints, cannot 'evolve' with time.

A possible route to the resolution of the puzzle is through the introduction of a 'clock variable' and conditional probabilities(see for instance\cite{Rovelli:1990jm}).
The idea is to promote one of the degrees of freedom to a clock variable. The rest of the observables can now be thought to evolve in the clock time. Which is to say, one can construct observables $B_T$ defined as the value of the degree of freedom $B$ when the clock degree of freedom $T$ takes some value. In the quantum theory one obtains conditional probability amplitudes $\Psi (x_i,t)$, where $t$ is the value of the clock degree of freedom and $x_i$ are the other degrees of freedom. 

An example is the parametrized particle(see \cite{kchar2} for more details on this theory). One considers a Newtonian particle in say, 1 dimension. Then one promotes the Newtonian time parameter $T$ to a degree of freedom by introducing an auxiliary variable $\tau$. This gives us a theory with two degrees of freedom
$T(\tau)$ and $X(\tau)$. Of course, one of the degrees of freedom in the theory is fake, and this shows itself in the reparametrization invariance of $\tau$. In the Hamiltonian picture, this translates to the Hamiltonian becoming a constraint $\tau$. This is as one would expect, because the Hamiltonian generates translation in $\tau$, and $\tau$ is a gauge parameter. 

Now one can quantize this theory, and obtain (in principle) the physical Hilbert Space by solving the constraints. This would give one wave-functions $\psi (X,T)$. One can get back usual quantum mechanics simply by interpreting T as a clock variable. Which of course is the interpretation one uses in usual quantum mechanics. 

Minisuperspace models constitute another set of examples. Loop Quantum Cosmology, for instance, is an example of this (see \cite{Bojowald:2008zzb} for a review of LQC). There one quantizes a FRW universe coupled to a single scalar field. One obtains a physical Hilbert Space of probability amplitudes of the type $\psi(c,\phi)$ where $c$ is a degree of freedom related to the scale factor. Again interpreting the scalar $\phi$ as a clock, one can think of these amplitudes as giving conditional probabilities for $c$ when the scalar takes some particular value. 

In this essay we would like to point out that this clock variable/ conditional probability interpretation suffers from a couple of conceptual problems that to the best of our knowledge have not been discussed in the past. In our opinion, these problems render it unsatisfactory. 

The first problem has to do with the question of simultaneous measurements. The second problem has to do with measurement postulate for the clock. We elaborate on these issues in the next two sections. 

\section{The question of when} 

In this section we consider the problem one runs into when one wishes to interpret a wave-function obtained as above as the 'wave-function of the universe'. To be sure, there is the usual problem of quantum mechanics of cosmology (or more generally, closed systems) - who observes the universe? But even before we get to the observer there is another conceptual problem.

Recall the interpretation of conditional probabilities: 'The probability of observing some $X$ \textit{when} the clock reads $T$'. The problem is in that 'when' - what does it mean? In this game the only definition of time is as the measurement result of the clock variable, there is no external clock $T'$ to tell us that the measurement of the clock degree of freedom $T$ and some other degree of freedom $X$. There is therefore no meaning that can be attached to that 'when'. 

One could argue that there could be some synchronizing operation one could use to define the simultaneous measurement of the two degrees of freedom. However this runs into the problem of closed systems mentioned above, the wave-function already includes the entire universe and there is nothing left to use to define a synchronization. 

However one might take the whole 'wave-function of the universe' thing somewhat less seriously and allow for a classical observer as well as such a synchronization process - which is to say one has an observer, a clock, some light beams to be used for synchronization and the rest of the universe. However this would lead to the second problem we discuss in this essay.

\section{What is the measurement postulate for the clock?}

One might think that allowing for an observer, a clock and a synchronization mechanism would solve all our problems. Because after all, does it not make the measurement procedure identical to what we do in usual experiments? 

While measuring some observable $\hat{A}(t)$ we do after all measure the time on a clock, measure the observable itself and then correlate the two. How is that any different from what one would do in this case?

The difference of course lies in the fact that, in the case of usual quantum mechanics, the clock was a classical observable. Measuring it (which in this case is also same as using light beams to synchronize its measurement with that of the other observable) simply had no significant effect on the clock. 

Indeed in an operational description, the classical clock can be regarded as an independent classical degree of freedom, evolving under some Hamiltonian. This is why, despite mathematical equivalence, the quantum theory of the parametrized particle is inequivalent to the usual quantum mechanics of a particle. Because in quantum mechanics, every measurable quantum degree of freedom is associated with a measurement postulate, which tells us what happens when we make immediate further measurements of the same degree of freedom.

The clock is exempt from this postulate in ordinary quantum mechanics, because the clock is a classical degree of freedom. The clock in a parametrized theory or any other such theory cannot be exempt from a measurement postulate, because that would mean that such a postulate is selectively applied to some degrees of freedom\footnote{Of course, the real world clock is also a quantum degree of freedom and this indeed creates problems in the quantum realm\cite{peres}. The important difference is that it is an unconstrained one, evolving according to its own Hamiltonian and therefore has an independent wave function. This means that the problems described in the next paragraph do not apply to it. }.

One may counter with the claim that the measurement postulate is only to apply to the wave-function and not to the partial observables in themselves. But the problem with this counter is that it makes the measurement of the wave-function ill defined. This is because now the measurement is complete only after two different partial observables have been observed, both of which are quantum degrees of freedom. And this brings back problems related to the definition of measurement. If (even allowing for some synchronization procedure) the clock and the other degree of freedom are measured out of sync, does it still constitute a measurement? Clearly such a measurement cannot be interpreted in terms of conditional probability amplitudes, simply because we did not keep tab on the clock. But if only simultaneous measurements are to be considered as measurements it leads to absurdity: how does the state of the system change (or not) based on whether its measurement was synchronized with the measurement of a different system?

As we can see, one opens a Pandora's box of paradoxes and inconsistencies if one tries to cast the clock as a constrained quantum degree of freedom and asks for the corresponding measurement postulate.

\section{Conclusion}

Conditional probabilities and relational observables are central to most canonical approaches to quantization of gravity and are widely used for minisuperspace models. However, we showed that this ideas run into several conceptual difficulties if issues related to their measurement are taken seriously. This poses severe interpretational problems for the canonical approach to quantum gravity.


\begin{thebibliography}{50}
\bibitem{kuchar}
 K.~Kuchar, 
 \textit{Conceptual Problems of Quantum Gravity}, ed. A. Ashtekar and J. Stachel (Birkhauser,1991)

\bibitem{Rovelli:1990jm} 
  C.~Rovelli,
  Phys.\ Rev.\ D {\bf 42}, 2638 (1990).
  doi:10.1103/PhysRevD.42.2638
  
\bibitem{kchar2}
 K.~Kuchar,
 \textit{Proceedings of the 4th Canadian Conference on General Relativity and Relativistic Astrophysics}, ed. G.
Kunstatter et. al. (World Scientific, New Jersey 1992)

\bibitem{Bojowald:2008zzb} 
  M.~Bojowald,
  Living Rev.\ Rel.\  {\bf 11}, 4 (2008).
  
\bibitem{peres}
A.~Peres,
Am.\ J.\ Phys {\bf 48}, 552 (1980) 

\end{thebibliography}
\end{document}